\documentstyle[prl,twocolumn,aps]{revtex}

\begin{document}
\draft

\title{Bound entanglement and continuous variables}

\author{Pawe\l{} Horodecki$^{1,2,}$\cite{poczta1} and
Maciej Lewenstein$^{2,}$\cite{poczta2}
}
\address{
$^1$ Faculty of Applied Physics and Mathematics,
Technical University of Gda\'nsk, 80--952 Gda\'nsk, Poland \\ 
$^2$ Institute f\"ur Theoretische  Physik,
Universit\"at Hannover, D-30167 Hannover, Germany} 
\maketitle
\begin{abstract}
We introduce the definition
of generic bound entanglement for the case of continuous variables.
We provide some examples of bound entangled states
for that case,  and discuss their
physical sense in the context of quantum optics.
We rise the question of whether the  entanglement of these states
is generic. As a byproduct we obtain a new many parameter family of
bound entangled states with positive partial transpose. 
We also point out that the ``entanglement witnesses''
and positive maps revealing the corresponding bound entanglement
can be easily constructed.
\end{abstract}
  
\narrowtext

Entanglement is a fascinating property of quantum states  evoking fundamental 
\cite{EPR,Sch},  as well as practical questions.
In the context of information theory, it has been proven to be useful 
in quantum cryptography \cite{Ekert}, 
quantum dense coding \cite{geste}, quantum teleportation
\cite{Tel} and quantum computation\cite{shor}.
In order 
to make entanglement useful despite of  the noise coming from 
interactions with environment, the idea of 
noisy entanglement distillation has been introduced \cite{Dist}
The distillation 
problem, i.e. the question which states are distillable, 
has a simple solution for low dimensional quantum systems:
two spin-${1 \over 2}$
particles, or
spin-1 plus spin-${1 \over 2}$ systems \cite{pur}. In those cases 
{\it any } 
noisy entanglement can be distilled to
maximally entangled form:
For larger spins the existence bound entanglement (BE) 
i. e. entanglement which is not distillable has been demonstrated
\cite{bound}.  
BE represents the result of nontrivial irreversible process
in which entanglement is confined to the physical system.
It was shown that there is connection of BE with other very interesting
quantum phenomena, called nonlocality without entanglement \cite{nonlocality}
(see \cite{epr} for discussion).

It is not trivial to provide examples of states which are 
BE. It has been shown \cite{bound}
that any state which is entangled and at the same time 
satisfies positive partial transpose (PPT) condition 
\cite{Peres} is bound entangled. 
The existence of PPT entangled states
was discussed in \cite{sep} and the first 
explicit examples were provided in \cite{tran}.
The first systematic procedure of constructing such states, 
employing unextendible product basis (UPB)   was 
provided in Refs. \cite{UPB1}. 
In the mathematical literature 
the first examples of matrices which can be treated as prototypes 
of PPT entangled states were provided by 
Choi  \cite{Choi}.
Here, we shall use the generalised structure of the Choi matrices
to provide the first examples of PPT entangled states
for continuous variables. 

Let us recall that the PPT separability condition \cite{Peres}, 
applied to 
a density matrix
$\varrho$ requires that the partial transposed  
matrix $\varrho^{T_B}$ is still a legitimate state.  
The matrix $\varrho^{T_B}$
associated with an  arbitrary product orthonormal $|i, j\rangle$
basis is defined as:
\begin{equation}
\varrho^{T_B}_{m\mu,n\nu}\equiv
\langle m, \mu| \varrho^{T_B}| n,\nu \rangle=
\varrho_{m\nu,n\mu},
\end{equation}
and the Peres criterion \cite{Peres} requires that  $\varrho^{T_B}\ge 0$
for separable $\varrho$.  
This statement is {\it valid} also for 
the cases when the state is defined on 
infinite dimensional Hilbert space.

Although the existence 
of BE states for finite dimensions has been proved,
it has not been known so far whether nontrivial
examples of BE states exist  in the infinite dimensional case.
In fact, main investigations of entanglement in the continuous
variables area were performed for pure states
resulting in  nonlocality effects \cite{Konrad}, new versions of   
teleportation \cite{Br1}, quantum computation \cite{Br2}, quantum error
correction \cite{Br3}
and quantum dense coding \cite{Br4}.
For mixed states, however, the PPT condition for continuous variables
has been, so far, analysed only in situations,   in which 
 it is necessary and sufficient
for separability. In particular, it  
has been  shown this is the case 
for Gaussian states \cite{Durr,Simon}.

In this paper 
we discuss bound entanglement for continuous variables.
We define the requirement any generic 
bound entangled state must satisfy in that case.
We provide the first examples of nontrivial PPT entangled states, 
{\it ergo} BE states
for continuous variables.
We rise the question how generic they are,  
and  discuss also the problem of physical
realization of such states.

Of course, one can simply construct  a  trivial example.
Consider, say $3 \otimes 3$ BE state $\sigma$,
and the infinitely dimensional Hilbert space
${\cal H}´$. Let us define 
infinitely many ``copies'' of $\sigma$
labelled by $\sigma_n$, each of 
which has the matrix elements of the original $\sigma$,
but in basis $S_n=\{ |i,j\rangle \}_{i,j=3n}^{3n+3} $.
Let $\{ p_i \}_{i=1}^{\infty}$ be a infinite sequence  of nonzero 
probabilities, $\sum_{i=1}^{\infty} p_i=1$.
Then the following state
\begin{equation}
\tilde{\sigma}=\mathop{\oplus}\limits_{n=1}^{\infty}p_n \sigma_n,
\label{oszuk}
\end{equation}
is bound entangled, 
but it has a trivial form from the continuous variables point of 
view\cite{simple}.
Actually, it can be reproduced with arbitrary accuracy performing
local transformations on states
which are of the $3 \otimes 3$ type.
Moreover,  they can be produced in a {\it reversible way}.
This follows from the fact that 
the states $\sigma_n$ and $\sigma_{n'}$
are {\it locally} orthogonal \cite{locort}. That means that Alice and Bob
can distinguish them using  local  quantum actions and classical communication
(LQCC) only. 
This  can be done in a reversible way as both persons
can forget the results of measurements.
In effect,  there is no entanglement between states
belonging to sets of Alice vectors $|i\rangle_{i,j=3n}^{3n+3}$ and
Bob ones $|i\rangle_{i,j=3n'}^{3n'+3}$
for $n\neq n'$.

Thus, in the case of the states 
(\ref{oszuk}) we deal effectively with $3 \otimes 3 $ 
type entanglement only. 
What does that  mean from the formal,  and more rigorous point of view?
One should ask first what does that mean that 
a state represents a {\it generic}  $N \otimes N$
type entanglement.
The answer to this question can be obtained  immediately  using the 
recently introduced definition of Schmidt rank
\cite{srank} for mixed states.
Let us recall the definition: 

{\bf Definition.-} Bipartite density matrix  $\varrho$ 
has Schmidt rank $K$ iff 
(i) for any decomposition of $\rho$, $\{p_i \geq 0, |\psi_i \rangle\}$ with
$\rho=\sum_i p_i |\psi_i \rangle \langle \psi_i|$ at least one of vectors
$| \psi_i \rangle$ has Schmidt rank $K$,  and (ii) there exists a
decomposition
of $\varrho$ with all vectors $\{ | \psi_i \rangle \}$ of Schmidt rank at most
$K$.

Thus it is  natural from the physical 
point of view to say that the state represents generic rank
$K$ entanglement iff it has Schmidt rank $K$. 
We introduce therefore:

{\bf  Definition.-}
A state $\varrho$ represents generic {\it continuous 
variables} or {\it infinite Schmidt rank entanglement} iff 
it is  the limit of   states $\varrho_n$
of Schmidt rank $K_n$, with $\lim_n K_n=\infty$.

In the following  we shall  focus on the question of existence
of  {\it generic} 
continuous variables entanglement, which would be at the same time bound
entanglement, i.e. entanglement which cannot be 
distilled. We will construct PPT 
states for continuous variables and argue that 
they represent generic infinite Schmidt rank entanglement.

For this aim consider first the state 
\begin{equation}
|\Psi \rangle=
\sum_{n=1}^{\infty} a_n |n,n \rangle,
||\Psi||^2=\sum_{n=1}^{\infty} |a_n|^2=q< \infty,
\label{przyklad}
\end{equation}
and the family of states
\begin{equation}
|\Psi_{mn} \rangle=
c_m a_n |n,m \rangle 
+(c_m)^{-1} a_m |m,n \rangle, 
\label{cztery}
\end{equation}
for $n<m$ with (in general) complex $a_n$ and $c_n$ such that
$0<|c_{n+1}|<|c_n|<1.$
Let us assume that the sum 
$\sum_{n=1}^{\infty}
\sum_{m>n}^{\infty} ||\Psi_{mn} ||^2 $ is finite. 
This can be achieved for example by setting 
$a_n=a^{n}$, $c_n=c^{n}$ for some $0<a<c<1$.
The sum becomes then a double geometric series and is given by
$a^4c^4(1-c^2)^{-1}(1-a^2c^2)^{-1} +
a^6(c^2-a^2)^{-1}(c^2-a^4)^{-1}.$
Under the above assumptions the  matrix 
\begin{equation}
\varrho=\frac{1}{A}(|\Psi \rangle \langle \Psi | +
\sum_{n=1}^{\infty}
\sum_{m>n}^{\infty} |\Psi_{mn} \rangle \langle \Psi_{mn}|),
\label{stan}
\end{equation} 
with the normalising factor 
$$A\equiv ||\Psi||^2+\sum_{n=1}^{\infty}
\sum_{m>n}^{\infty} ||\Psi_{mn} ||^2,$$
represents a legitimate quantum state 
in the space $l^2({\cal C}) \otimes l^2({\cal C})$,
where $l^2({\cal C})$ is the space of all complex 
sequences $\{ z_n \}$, $\sum_{n=1}^{\infty} |z_n|^2 < \infty$.
It can be seen by inspection that the above state satisfies 
$\varrho=\varrho^{T_B}$, and thus 
has the
PPT property. In fact 
we have chosen  $|\Psi_{mn} \rangle$ in Eq. (\ref{cztery})
to ensure this property. It follows immediately that pure state entanglement 
cannot be distilled from (\ref{stan}). For this aim simple arguments 
from the  Ref.
\cite{bound} can be  recalled, and applied  to the
separable superoperators in infinitely dimensional space.

Subsequently we shall show that the above states are entangled, and, thus 
being the PPT states represent bound entanglement.
To this aim we shall prove that the state (\ref{stan}) 
has the following property:

{\bf Property.-} Any local measurement of state $\varrho$
\begin{equation}
\varrho \rightarrow \varrho'={P \otimes Q \varrho P \otimes Q  \over Tr(
P \otimes 
Q \varrho P \otimes Q) }
\end{equation}
by means of 
$P$, $Q$  projecting onto the space ${\rm span} \{ |n_1 \rangle, ..., |n_{K} 
\rangle \} $ on Alice and Bob's  sides respectively 
results in  $K \otimes K$ bound entangled state $\varrho'$.
From the above property it follows immediately that
$\varrho$ is a bound entangled state.

{\bf Proof.-} Let us first prove that for any $K$ the state $\varrho'$ is a 
$K\otimes K$ BE state. 
To this aim we observe that after local filter operation 
corresponding to the operator $V={\rm diag}[a^{-1}_{n_1}, 
..., a^{-1}_{n_{K}}]$ on the Alice side,
and a suitable unitary transformation $U_1 \otimes U_2$ 
(that transforms $|n_m\rangle\to|m\rangle$ on both 
Alice and Bob's sides), 
the state becomes proportional to the particularly simple matrix:
\begin{equation}
\Sigma \equiv
|\Phi \rangle \langle \Phi | +
\sum_{n=1}^{K} 
\sum_{m>n}^{K} |\Phi_{mn} \rangle \langle \Phi_{mn} |,
\label{s}
\end{equation}
with 
$|\Phi \rangle=
\sum_{n=1}^{K} |n,n \rangle$, 
and 
$|\Phi_{mn} \rangle=
{\alpha_m} |n,m \rangle
+ {\alpha_m}^{-1}  |m,n \rangle$.
In the following, we shall use the general definition 
of $\varrho'$ as well, 
but after the above defined local action the state $\Sigma$ is proportional 
to the matrix with parameters $\alpha_m=c_{n_m}$.
It means in particular that $0<\alpha_{i+1}<\alpha_{i}<1$.
We shall prove  that  $\Sigma$ does not have any product vector in 
its range, {\it ergo} that, if normalised, $\Sigma$ is an entangled state.
That will mean, however, also that the state $\varrho'$
is bound entangled, since the local filtering and 
the local unitary operations are reversible with nonzero probability.

Suppose, that there was a nonzero product state $|\psi, \phi \rangle$ in the range 
of the matrix (\ref{s}). Since the range of $\Sigma$ is spanned by its
eigenvectors, there would exist some $g$, $g_{ij}$,
$i=1,...,K, j>i$ such that:
\begin{equation}
g|\Phi \rangle  +   
\sum_{i=1}^{K}
\sum_{j>i}^{K} g_{ij}|\Phi_{ij} \rangle=
|\psi, \phi \rangle. 
\label{rownosc}
\end{equation}
Suppose first that we would have $g\ne 0$ in (\ref{rownosc}).
Then, we could set $g=1$, and  the following constraints 
would 
immediately follow from (\ref{rownosc}): 
\begin{eqnarray}
&&\psi=[x_1,...,x_K], \nonumber \\
&&\phi=[(x_1)^{-1},...,(x_K)^{-1}]
\label{iksy}
\end{eqnarray}
for some numbers $\{ x_i \}$ which are all nonzero.
Substituting (\ref{iksy}) into (\ref{rownosc})
leads to the equations:
\begin{eqnarray}              
&&g_{ij} \alpha_j={x_i \over x_j}, \nonumber \\
&&g_{ij} (\alpha_j)^{-1}=\left({x_i \over x_j}\right)^{-1}, \nonumber \\
&&{\rm for}\ i=1, ..., K, \ j>i. 
\label{gie}
\end{eqnarray}
Since numbers
$\{ x_i \}$ are nonzero, the coefficients
$\{ g_{ij} \}$ have to be nonzero too.
Thus we have  $\alpha_j^2=({x_i \over x_j})^2$
for every $i=1, ..., K-1, \ j>i$.
We can, however, put $x_1=1$, and then we get that all $x_i^2$'s  are equal,
and that 
\begin{equation}
\alpha_j^2=1, \ \mbox{for} \ j=2, ..., K-1. 
\label{conditions}
\end{equation}
This is in contradiction with 
the condition 
$0<\alpha_{i+1}<\alpha_{i}<1$ fulfilled by $\Sigma$.

Consider now the case when $g=0$ holds in equation 
(\ref{rownosc}). That would mean, 
keeping the same notation for  
$\psi$, i.e. $|\psi \rangle=[x_1,...,x_K]$  
that we could have 
$|\phi \rangle=[y_1,...,y_K]$  with $y_i\neq 0$ iff $x_i = 0$.
But, if we examine the equation (\ref{rownosc})
under those conditions 
we get immediately that all $g_{ij}$ parameters must
vanish, so that  the whole LHS of 
the equation becomes then equal to zero. 
It means that there is no 
product vector in the range of the matrix 
$\Sigma$. 
Following previous discussion it is not difficult 
to see that the same holds for states 
$\varrho'$, which are thus (by virtue of the range criterion
of Ref. \cite{tran}) entangled. 
Collecting all the above observations,
we see that the {\bf Property} of 
the original matrix $\varrho$ holds. $\Box$
 
Unfortunately, it is  not easy to see that $\varrho'$ ($\varrho$) represents
the generic rank $K$ ($\infty$) entanglement. In fact
$\varrho'$ contains in the mixture the pure state of Schmidt rank $K$
which cannot be  distinguished from
the rest of the mixture in a reversible manner,  since its reduced density 
matrix has full rank $K$. Certainly, $\varrho$  does not consist of locally
orthogonal representation of finite Schmidt rank entanglement.
This can bee easily seen from the fact that local orthogonality
is a stronger property than orthogonality in case 
of pure states vectors.
Had $\varrho$ been a locally  orthogonal mixture of finite Schmidt rank
state, the 
eigenvectors of $\varrho$ would have been locally orthogonal, 
and of finite Schmidt rank, which is obviously not true, since  
one of the eigenvectors 
 of $\varrho$ is of infinite Schmidt rank. Nevertheless,  
we have not been able  to show, so far,  that the Schmidt rank 
of the proposed states 
is infinite. It is, however, quite likely  
that  either these states, or some modification of them
posses that property. 

{\it Finite dimensional bound entangled states .-} 
It is remarkable that as a byproduct we have obtained here 
a new family of $K \otimes K$ bound entangled states 
for an arbitrary $K$.
These are the states 
$\sigma={ \Sigma \over Tr(\Sigma)}$,
with $\Sigma$ violating  one (or more) of the 
$K-1$ conditions (\ref{conditions}).
In this notation we recall the Choi matrix
as a special case of $\Sigma$ with $K=3$, and all $\alpha$'s 
equal to 2 (see \cite{Choi}). 

{\it The corresponding ``entanglement witnesses''
and positive maps .-}
It should be pointed out that any BE state 
from the last paragraph (i. e. $\frac{\Sigma}{Tr\Sigma}$ 
violating condition (\ref{conditions}))
has no product vector in its range. 
Thus the projector $P$ onto its range has no 
product vector in its range as well. 
This is the same as in the projector 
orthogonal to UPB set of vectors \cite{UPB1}, 
and thus {\it mutatis mutandis} the approach 
from the paper \cite{Barbara} can be immediately 
applied to reproduce both entanglement witnesses,
as well as the corresponding positive maps.

{\it Possibility of physical realization.-} 
Let us now discuss a possibility of physical realization 
of the states of the type of $\varrho$, as states of two photon 
modes of electromagnetic field of equal 
or similar frequency, and  orthogonal
polarisations.  

Let us set $a_n=e^{-\beta n}$, $c_n=e^{-\gamma n}$, 
$\gamma<\beta$, and let us denote the 
corresponding photon creation and annihilation
operators of the two modes considered as 
$A^\dagger$, $A$, $B^\dagger$, $B$, respectively. 
The state
$\varrho$ can be represented as a mixture 
\begin{equation}
\varrho \sim |\Psi \rangle
\langle \Psi|+
\sum_{k=1}^{\infty} \varrho_k,
\label{dup}
\end{equation}
 where 
\begin{equation}
\varrho_k = V\delta(B^\dagger B - A^\dagger A -k)V^{\dag},  
\label{rhok}
\end{equation} 
where $V=e^{-\beta A^\dagger A - \gamma B^\dagger B} + U e^{-(\beta
-\gamma) B^\dagger B}$, 
$U$ is the unitary operator that transforms $A$ photons into $B$ photons,
while the operator function $\delta(.)$ is the 
operator valued Kronecker delta, 
$\delta(x)=0$, except for $x=0$,
when $\delta(x)=1$. 
We propose the following prescription in order to 
generate the states corresponding to the subsequent terms 
in the mixture (\ref{dup}):  

\noindent
i) The state $|\Psi \rangle \sim \exp(-
\gamma B^\dagger A) |0,0\rangle$  can be created 
as a two mode squeezed state, for instance in the
process of degenerate  parametric amplification
\cite{Konrad,Br4}. In fact, such states have been used for  
teleportation  with continuous variables \cite{Br4}.

\noindent ii) Each of the terms $\rho_k$ can be 
obtained by applying the positive operator valued measurement 
to the states $\delta_k=\delta(B^\dagger B - A^\dagger A -k)$, that transforms 
(with some probability)
$$\delta(B^\dagger B - A^\dagger A -k)\to V\delta(B^\dagger B - 
A^\dagger A -k)V^{\dag}.$$

\noindent iii) The operator $V$ can be realized by using an ancilla (say a two
level atom medium) with levels $|0\rangle$ and $|1\rangle$ that undergoes
 losses (via spontaneous emission, or ionization)
proportional to $-\beta A^\dagger A - \gamma B^\dagger B$ for the level
$|0\rangle$ and $-(\beta
-\gamma) B^\dagger B$ for $|1\rangle$, and dispersive dynamics governed by the
Hamiltonian $h\propto |0\rangle\langle 0| +(A^\dagger B + B^\dagger
A)|1\rangle\langle 1|$. One should first prepare the ancilla in the state
$(|0\rangle+|1\rangle$, wait appropriate  time
so that the dispersive dynamics
will realize the controlled $U$ operation, 
$|0\rangle\langle 0| +U|1\rangle\langle
1|$, and project the system then onto the initial  ancilla state.

\noindent iv) The main problem 
consist thus in generating the states $\delta_k$, 
$$\sum_n\sum_{k=1}^{\infty}|n,n+k\rangle\langle n,n+k|.$$ 
These states cannot be
normalised, but we can always regularize them by 
including part of the thermal noise operators into their definition. In order
to realize the states $\delta_k$, we propose to use a $K$ level ancilla ($K$-level
atom) constituting a Kerr medium with the states $|i\rangle$, $i=1,\ldots,K$. The
Hamiltonian should now be 
$\tilde h=\sum_i \Delta_i(A^\dagger A, B^\dagger B)
|i\rangle\langle i|$, where the intensity dependent energy shift of the $i$-th level
is 
$\Delta_i(A^\dagger A, B^\dagger B)=x_i(A^\dagger A-B^\dagger B+k)$.
 Kerr effect
should be here of the opposite sign for the two modes in question, 
but of the same magnitude. 
The last term in $\Delta_i$ represents normal linear phase shift. 

\noindent v) The idea is then to prepare the $K$-level 
ancilla in a more or less equal
weight superposition of the states $|i\rangle$, evolve then the system
according to the dynamics governed by $\tilde h$, and project
at the end on the initial state of the ancilla.
Note that in such case, the action of such "phase shifter" on the state
$|n,m\rangle$ will be  
\begin{equation}
|n,m\rangle \rightarrow \sum_{i} e^{ix_i(n-m+k)}|n,m\rangle,
\end{equation}
which, if $K$ is sufficiently large and $x_i$ cover a broad variety of phases,
provides  a finite bandwidth
approximation of the Kronecker delta.

Summarising, we have presented the first nontrivial example of 
bound entangled states
for continuous variables. We have presented strong evidence that these state
represent  generic bound state entanglement of infinite rank.  

PH acknowledges support
from Deutscher Akademischer Austauschdienst and 
partial support from Polish Committee for Scientific Research,
grant No. 2 PO3B 103 16. This work has been also supported by the
DFG (SFB 407, Schwerpunkt "Quanteninformationsverarbeitung"),
and European Union IST Programme EQUIP.


\begin{references} 

\bibitem[*]{poczta1}
E-mail address: pawel@mifgate.pg.gda.pl

\bibitem[**]{poczta2}
E-mail address: lewen@itp.uni-hannover.de 

\bibitem{EPR}
A. Einstein, B. Podolsky  and N. Rosen, Phys. Rev. {\bf 47}, 777 (1935).

\bibitem{Sch}
E. Schr\"odinger, Proc. Cambridge Philos. Soc. {\bf 31},  555 (1935).

\bibitem{Ekert}
A. Ekert, Phys. Rev. Lett. {\bf 67}, 661 (1991).

\bibitem{geste}
C. H. Bennett  and S. J. Wiesner, Phys. Rev. Lett. {\bf 69}, 2881 (1992).

\bibitem{Tel}
C. Bennett, G. Brassard, C. Crepeau, R. Jozsa, A. Peres and W. K. Wootters,
Phys. Rev. Lett.  {\bf 70}, 1895  (1993); 
for experiments see 
D. Bouwmeester, J.-W. Pan, K. Mattle, M. Elbl, H. Weinfurter and
A. Zeilinger, Nature (London) {\bf 390}, 575 (1997);
D. Boschi, S. Brance, F. De Martini, L. Hardy and S. Popescu,
Phys. Rev. Lett. {\bf 80}, 1121 (1998).


\bibitem{shor} cf. P. W. Shor, quant-ph/0005003

\bibitem{Dist}
C. H. Bennett, G. Brassard, S. Popescu, B. Schumacher, J. Smolin and
W. K. Wootters, Phys. Rev. Lett. {\bf 76}, 722 (1996);
D. Deutsch, A. Ekert, R. Jozsa, Ch. Macchiavello, 
S. Popescu and A. Sanpera, Phys. Rev. Lett.  {\bf 77},
2818 (1996).

\bibitem{pur}
M. Horodecki, P. Horodecki and R. Horodecki,
Phys. Rev. Lett. {\bf 78}, 574 (1997).

\bibitem{bound}
M. Horodecki, P. Horodecki and R. Horodecki, Phys. Rev. Lett.
{\bf 80}, 5239 (1998);
P. Horodecki, M. Horodecki and R. Horodecki.
Phys. Rev. Lett. {\bf 82}, 1056 (1999).

\bibitem{nonlocality}
C. H. Bennett, D. DiVincenzo, Ch. Fuchs, T. Mor, E. Rains, P. Shor, J.
Smolin, and
W. K. Wootters,  Phys. Rev. A {\bf 59}, 1070 (1999).

\bibitem{epr}
R. Horodecki, M. Horodecki, and P. Horodecki, Phys. Rev.
A, {\bf 60}, 4144 (1999). 

\bibitem{Peres} A. Peres, Phys. Rev. Lett. {\bf 76}, 1413 (1996).

\bibitem{sep}
M. Horodecki, P. Horodecki and R. Horodecki, Phys. Lett. A{\bf 1},  223 (1996).

\bibitem{tran} P. Horodecki, Phys. Lett. A, {\bf 232}, 233 (1997). 

\bibitem{UPB1}
C. H. Bennett, D. DiVincenzo, T. Mor, P. Shor, J. Smolin, and B. M. Terhal,
 Phys. Rev. Lett. {\bf 82}, 5385 (1999);
D. DiVincenzo, T. Mor, P. Shor, J. Smolin, and B. M. Terhal, 
quant-ph/9908070;  D. Bru\ss\ and A. Peres, Phys. Rev. A {\bf 61} 30301(R)
(2000).

\bibitem{Choi} M. D. Choi, Proc. Sympos. Pure Math {\bf 38}, 583 (1982). 

\bibitem{locort}P. Horodecki, M. Horodecki and R. Horodecki, Acta Phys.
Slov. {\bf 48}, 141 (1998).

\bibitem{Konrad}K. Banaszek, K. W\'odkiewicz, Phys. Rev. Lett. {\bf 82},
2009, (1999).

\bibitem{Br1}P. van Loock and  S. Braunstein, Phys. Rev. A,
{\bf 61} 010302(R) and
references therein.

\bibitem{Br2}S. Braunstein, Phys. Rev. Lett. {\bf 80}, 4084 (1998).

\bibitem{Br3}S. Lloyd and S. Braunstein, Phys. Rev. Lett. {\bf 82}, 1784 (1999).

\bibitem{Br4}S. Braunstein and  J. Kimble, quant-ph/9910010, and references
therein; for experiments see A. Furusawa {\it et al.} Science {\bf 282}, 706
(1998). 

\bibitem{Durr} L. M. Duan, G. Giedke, J. I. Cirac, and P. Zoller,
Phys. Rev. Lett.  {\bf 84} 2722 (2000).

\bibitem{Simon} R. Simon, Phys. Rev. Lett.  {\bf 84} 2726 (2000).


\bibitem{simple} This is the most simple, and  to some extend typical 
example of a nongeneric BE state.

\bibitem{srank} B. Terhal and P. Horodecki, Phys. rev. A {\bf 61}
040201(R) (2000).

\bibitem{Barbara}B. Terhal,  quant-ph/9911057. 

\end{references}
\end{document}